\documentclass[showpacs,twocolumn,pre]{revtex4}
\usepackage{psfrag,epsfig,amsfonts,amssymb,amsmath}

\newcommand{\ord}{{\cal O}}

\begin{document}
\title{Paradoxical non-linear response of a Brownian particle}
\author{Ralf Eichhorn}
\author{Peter Reimann}
\affiliation{Universit\"at Bielefeld, Fakult\"at f\"ur Physik, 33615 Bielefeld, Germany}
\date{\today}
\begin{abstract}
We consider a Brownian particle in a ``meandering''
periodic potential when the ambient temperature is a
periodically or stochastically varying function of time.
Though far from equilibrium, the linear response of
the particle to an external static force is exactly the same as 
in the equilibrium case, i.e. for constant temperature.
Even more surprising is the non-linear response:
the particle slows down and then even starts to move in the 
direction opposite to the applied force.
\end{abstract}


\maketitle

The second law of thermodynamics requires that
the response of a system at thermal equilibrium 
to a static external force is a motion 
in the direction of that force.
In this Letter we address the geometrically 
constrained motion of a Brownian particle 
with the remarkable property that, 
though being far from
equilibrium, its linear response to
a static force $F$ is exactly the same as
in the equilibrium case.
In particular, the system is at rest in the absence
of the external perturbation $F$ and
for sufficiently small (positive or negative) 
$F$ it moves, as naively expected, in 
the same direction as the static force.
An even more remarkable, genuine non-equilibrium 
behavior arises in the non-linear response regime:
upon increasing $F$ to moderately large
(positive or negative) values, 
the particle slows down and then even starts to move in the 
direction opposite to the applied force $F$ !
Note that such a response behavior is
fundamentally different from so-called 
absolute negative conductance or mobility \cite{anm},
characterized by linear and non-linear response 
properties that are just reciprocal to those described above,
and from differential negative mobility \cite{dnm}, 
where the transport speed decreases with increasing 
force but never points in the direction opposite to 
that force.
Accordingly, also the underlying physics and, in particular, 
the considered systems are entirely different.
Furthermore, our present case has
nothing to do with a so-called ratchet effect \cite{reimann02},
where, by definition, the velocity of the Brownian particle 
is finite even in the absence of an
external force, thus keeping the same sign in an entire 
interval of positive and negative forces $F$, and
moreover increasing monotonically as a function of $F$.

The above announced response of a Brownian particle to an
external force requires the interplay of two indispensable 
ingredients: a non-linear dynamics and a source of
disequilibrium.
Both will be chosen in a way that is conceptually very simple 
and experimentally easy to realize.
As for the non-linear dynamics,
we consider an overdamped Brownian particle, whose
motion is confined to a quasi-one-dimensional meandering 
path in the $x$-$y$-plane, and which is subjected to
an externally applied homogeneous force along the
$x$-axis, see Fig. 1.
Experimentally, such a meandering state space can be 
realized e.g. by means of light forces
\cite{exp1} or morphologically via lithographic etching 
methods \cite{exp2}, while the external force can be
generated e.g. by electrical or magnetic fields, or 
by gravitation.
The second main ingredient of our model is a source of
disequilibrium.
Since the heydays of Carnot's machines, one of the
simplest examples in this respect is a temperature 
that periodically switches between two different 
values, see Fig. 2a.
Apart from this ``minimal'' example we will also admit 
somewhat more general time dependent temperatures $T(t)$,
consisting of a ``background'' or ``floor'' temperature
$T_0$, superimposed by short, but violent temperature
``pulses'', ``spikes'', or ``outbursts'', occurring at
random times $t_i$ and extending over characteristic
time intervals $\tau_i$, see Fig. 2b. 
In analogy to the periodic case,
the average interspike distance is denoted by
\begin{equation}
\tau:=\langle t_{i+1}-t_i\rangle \ ,
\label{eq1b}
\end{equation}
and the relative duration of the temperature spikes by
\begin{equation}
\vartheta:=\langle\tau_i\rangle/\tau \ .
\label{eq1c}
\end{equation}
Experimentally, such temperature variations
may be realized either directly via heating
e.g. with laser pulses  
or indirectly e.g. by pressure pulses.
We also note that such outbursts may not
necessarily be caused by thermal noise but may
as well be considered as a simplified model for sudden 
mechanical ``strikes'', ``quakes'', ``impacts'', 
``eruptions'' etc.
With a negligibly small ``background'' temperature $T_0$,
such a model may then describe a variety of experiments with
even quite large ``particles'' subject to some kind of
sporadic violent shaking.

The quantity of central interest is the 
average particle current along the $x$-axis
\begin{equation}
\langle \dot x\rangle := \lim_{t\to\infty}
\frac{x(t)}{t} \ .
\label{eq0}
\end{equation}
For ergodicity reasons, no ensemble average on the right 
hand side is needed and due to the long time limit, 
initial conditions do not matter.
For the symmetric example in Fig. 1b it is clear 
that the current $\langle\dot x\rangle$ is an odd function of
the applied force $F$ and in particular vanishes in the 
absence of the force.
For more general cases like in Fig. 1a, the former 
property is lost but, as we will show later, one still has
$\langle\dot x\rangle =0$ for $F=0$. 
In other words, our system does {\em not} exhibit a 
ratchet effect.

Parametrizing by $q$ the position of the particle
along its meandering route
(path length), its overdamped Brownian motion 
can be modeled by the Langevin-equation 
\cite{risken89,reimann02}
\begin{equation}
\label{eq1}
\eta \, \dot{q}(t) = -V'(q(t)) + \sqrt{2\eta \, k_{\mathrm{B}}T(t)} \; \xi(t) \, ,
\end{equation}
where $\eta$ denotes the damping coefficient, $k_{\mathrm{B}}$
Boltzmann's constant, and $\xi(t)$ a $\delta$-correlated,
unbiased Gaussian noise.
The force $-V'(q)$ in (\ref{eq1}) represents the projection 
of $\vec e_x F$ 
in Fig. 1 along the meandering path at the position parametrized 
by $q$. Thus $V(q)$ is proportional to $F$.
Denoting by $L_q$ the natural spatial period along the 
meandering path one can see from Fig. 1 that increasing the path length
$q$ by one period $L_q$ is tantamount to increasing the $x$-component 
of the particle position by one period $L_x$ and leaving the 
$y$-component unchanged.
As a first consequence, we can infer that $V(q+L_q)=V(q)-FL_x$ and hence
\begin{eqnarray}
\label{eq2}
V(q) & = & F\,[G(q) - q\, L_x/L_q]
\\
\label{eq3}
G(q+L_q) & = & G(q) \ .
\end{eqnarray}
As a second consequence, it follows that
the average velocity of actual interest (\ref{eq0})
is related to the average 
velocity $\langle\dot q\rangle := \lim_{t\to\infty}\{q(t)/t\}$
of the auxiliary dynamics (\ref{eq1}) according to
\begin{equation}
\langle\dot x\rangle = \langle\dot q\rangle \, L_x/L_q \ .
\label{eq9}
\end{equation}

Note that the periodic function $G(q)$ from (\ref{eq2}), (\ref{eq3})
is $F$-independent. In Fig. 3, we exemplify its explicit 
construction for the piecewise linear meandering path 
from Fig. 1b. Whence one can infer that
\begin{eqnarray}
G(q) & = & \left\{
  \begin{array}{l@{\hspace{3ex}}lccrcl}
    -q(1-L_x/L_q)         & \mathrm{for} & -l_2 & < & q & \leq & 0  \\
     q(\cos\varphi+L_x/L_q) & \mathrm{for} & 0    & < & q & \leq & l_1
  \end{array}
  \right.
\label{eq4}
\\[2ex]
G(q+ \!\! & l_1 & \!\! +l_2)  =  G(q)
\label{eq5}
\\[2ex]
L_x & = & 2\, (l_2 - l_1 \cos\varphi)
\label{eq6}
\\[2ex]
L_q & = & 2\, (l_1 + l_2) \ .
\label{eq7}
\end{eqnarray}

In order to study the linear response behavior of our system,
the deterministic force $-V'(q)$ in (\ref{eq1}), 
being $L_q$-periodic according to (\ref{eq2}), (\ref{eq3}),
is expanded into a Fourier-series.
Since each summand of this Fourier expansion is proportional 
to $F$ (cf. Eq. (\ref{eq2})) it follows that
in leading order $F$ (linear response) 
the effect in (\ref{eq1}) of each of those summands 
can be considered separately and their individual contributions 
to the current $\langle \dot q\rangle$ can simply be added up. 
For symmetry reasons one can infer \cite{reimann02} that any summand which 
exhibits harmonic oscillations as a function of $q$ does 
not lead to a systematic directed transport, 
i.e. does not contribute to $\langle \dot q\rangle$.
It remains the non-oscillating part of $-V'(q)$, which according
to (\ref{eq2}) is given by $F L_x/L_q$, and hence gives rise to
a current $\eta \langle \dot q\rangle = F L_x/L_q$ after
averaging out the noise in (\ref{eq1}).
Exploiting (\ref{eq9}) we thus can infer that
\begin{equation}
\langle\dot x\rangle = \left(\frac{L_x}{L_q}\right)^2\, \frac{F}{\eta} 
+ \ord(F^2) \ .
\label{eq13}
\end{equation}
The same result can also be derived in a mathematically more rigorous way
by means of a perturbation expansion in $F$ of the Fokker-Planck equation
equivalent to (\ref{eq1}). Moreover, it can be extended beyond the so far
considered case that $T(t)>0$ for all $t$.
The most remarkable property of the linear response behavior (\ref{eq13})
is its independence of the temperature $T(t)$: 
For strong temperature variations, i.e. {\em far from
equilibrium, it is exactly the same as in the equilibrium case} 
when the temperature is constant!

Next we turn to the non-linear response for not too small forces $F$.
The simplest case arises when $F$ is so large or
the ``background temperature'' $T_0$ is so small that in
the auxiliary dynamics (\ref{eq1})-(\ref{eq3}) the
effect of the fluctuations $\xi(t)$ is negligible between 
successive temperature spikes, cf. Fig. 2.
Furthermore, any temperature outburst is assumed
to be so violent, that the distribution of an ensemble of Brownian 
particles will be completely randomized, i.e. practically uniformly 
distributed within any interval comparable to the spatial period $L_q$. 
Moreover, during such an outburst, the particles will practically
not feel the details of the periodic part $G(q)$ of the
potential in (\ref{eq2}) but only its systematic ``tilt''
$-qFL_x/L_q$. 
Hence, the average displacement during an outburst of duration $\tau_i$ is
\begin{equation}
\Delta q_s(\tau_i)=\tau_i F L_x/(L_q\eta) \ .
\label{eq14}
\end{equation}
Between two outbursts, the particles start with a 
uniform initial probability distribution and then evolve deterministically
according to (\ref{eq1}) with $\xi(t)\equiv 0$. 
Focusing on the piecewise linear
example from Fig. 3b and Eqs. (\ref{eq4})-(\ref{eq7}) 
and on positive forces $F$, one readily finds that
the average net displacement during the 
time $\Delta t_i:=t_{i+1}-t_i-\tau_i$ 
between two temperature outbursts amounts to
\begin{eqnarray}
\!\!\!\!\!\!\!\!\!\!
\Delta q_0(\Delta t_i) \!\!& = &\!\!
[l_2^2-(l_2-v_2\Delta t_i)^2\Theta(l_2-v_2\Delta t_i)]/L_q
\nonumber\\
\!\!& - &\!\! [l_1^2-(l_1-v_1\Delta t_i)^2\Theta(l_1-v_1\Delta t_i)]/L_q
\label{eq15}
\end{eqnarray}
where $v_2:=|F|/\eta$ and 
$v_1:=|F|\cos\varphi/\eta$ are the two deterministic 
particle speeds associated with the two different slopes of the
piecewise linear potential $V(q)$ in Fig. 3b
and the Heaviside step functions $\Theta$
account for the fact that the particles stop to move once
they have reached a local potential minimum of $V(q)$.
The average displacement per temperature outburst
then follows as 
$\langle\Delta q_s(\tau_i)+\Delta q_0(\Delta t_i)\rangle$.
Dividing by the average interspike distance from (\ref{eq1b})
yields $\langle\dot q\rangle$. With 
(\ref{eq1c}), (\ref{eq9}), (\ref{eq14}) 
and taking into account that $\langle\dot x\rangle$ is
obviously an odd function of $F$, we finally obtain
\begin{equation}
\langle\dot x\rangle = F\, \frac{L_x}{L_q}\left[ 
\frac{\vartheta}{\eta}\, \frac{L_x}{L_q} + 
\frac{\langle\Delta q_0(t_{i+1}-t_i-\tau_i)\rangle}{|F|\, \tau}\right] \ .
\label{eq16}
\end{equation}
For random temperature bursts, the average 
$\langle\Delta q_0(t_{i+1}-t_i-\tau_i)\rangle$ has to be
evaluated according to their specific statistical properties.
In many cases, one expects that the result will be well approximated
by $\Delta q_0(\langle t_{i+1}-t_i-\tau_i \rangle)=
\Delta q_0([1-\vartheta]\tau)$, i.e.
\begin{equation}
\langle\dot x\rangle = F\, \frac{L_x}{L_q}\left[ 
\frac{\vartheta}{\eta}\, \frac{L_x}{L_q} + 
\frac{\Delta q_0([1-\vartheta]\tau)}{|F|\, \tau}\right] \ .
\label{eq17}
\end{equation}
For periodic bursts, Eqs. (\ref{eq16}) and (\ref{eq17}) are
even strictly equivalent.

For small $F$, one readily recovers the correct linear response 
behavior (\ref{eq13}) from (\ref{eq16}), though this small-$F$
regime was originally not included in our considerations above.
Next we turn to the case that $|F|$ is sufficiently large  
that the $\Theta$-functions in (\ref{eq16}) vanish.
Focusing on the case $l_1>l_2$ in Fig. 3, this 
amounts to the condition $(1-\vartheta)\tau |F| \cos\varphi/\eta>l_1$,
and (\ref{eq16}) with (\ref{eq7}), (\ref{eq15}) then takes the form
\begin{equation}
\langle\dot x\rangle = F\, 
\frac{L_x}{L_q}\left[\frac{\vartheta}{\eta}\, \frac{L_x}{L_q} -
\frac{l_1-l_2}{2\,|F|\, \tau}\right] \ .
\label{eq18}
\end{equation}
For sufficiently short relative pulse durations $\vartheta$ 
we thus expect the existence of an interval of
{\em moderately large $|F|$-values with the property that
the current $\langle\dot x\rangle$ will be positive for
negative $F$ and vice versa!}

In Figs. 4 and 5, results of numerical simulations are presented,
exhibiting a fairly good agreement with the approximation
(\ref{eq16}). In particular, both the predicted linear response 
and the paradoxical non-linear response behavior are confirmed.
While for the filled circles the assumptions made in the derivation
of the approximation (\ref{eq16}) are apparently quite well satisfied,
for the open circles there are notable deviations.
As mentioned above, for the periodic $T(t)$ in Fig. 4 the 
approximations (\ref{eq16}) and (\ref{eq17}) coincide, 
whereas for the stochastic $T(t)$ in Fig. 5 the more 
complicated expression (\ref{eq16}) is clearly superior.

Finally, we turn to a heuristic discussion of the general 
conditions under which the above paradoxical non-linear 
response behavior is expected.
Taking into account the obvious constraints
$0< \varphi < \pi/2$, $l_2/2 - l_1 \cos\varphi > 0$
in Fig. 3 and 
$l_1>l_2$, $(1-\vartheta)\tau |F| \cos\varphi/\eta>l_1$
[see above Eq. (\ref{eq18})] one can infer that
a necessary condition to get a current opposite to
$F$ in (\ref{eq18}) is $\vartheta<1/3$.
On the other hand, the shorter the relative duration 
$\vartheta$ of the temperature bursts in (\ref{eq18})
are, the more pronounced the effect will be.
In particular, for $\vartheta\to 0$ [i.e. $\delta$-spikes in Fig. 2]
the current $\langle\dot x\rangle$ in (\ref{eq18})
remains opposite to 
$F$ for arbitrarily large $|F|$.
So far we have assumed, that each temperature burst results
in an almost uniform particle randomization within
any period $L_q$. In the opposite case, i.e.
for not so violent bursts, the distribution will
remain notably peaked around the local potential minima of $V(q)$,
leading to a reduced displacement opposite to $F$ during the
subsequent relaxation period till the next burst.
In other words, the effect of interest diminishes with
decreasing intensity of the temperature bursts, and obviously
disappears altogether when the temperature variations in
Fig. 2 become negligible.
Similarly, one sees that our assumption of a negligibly small
``background'' temperature $T_0$, i.e. a deterministic relaxation dynamics
in between successive bursts, is optimal
in the sense that with increasing the noise intensity
$T_0$, the effect of interest decreases and finally disappears.
These predictions are confirmed by the open circles in Fig. 4,
whose deviations from the theoretical solid line can be traced 
back to the increased $T_0$-value for small $F$ and to the
reduced $T_{high}$-value for large $F$.
Finally, going over from the piecewise linear example in
Fig. 1b to the more general case in Fig. 1a leads to an 
increased deterministic relaxation time towards the local
potential minima of the associated auxiliary potentials 
$V(q)$ and hence again to a reduction of the effect of interest.
Notwithstanding, it is clear that a current opposite to an applied
force of suitable magnitude will arise whenever the outbursts
are sufficiently short and violent, $T_0$ is sufficiently small,
and if the meandering path is such that a motion
along this path with constant velocity
will result in a back-and-forth motion when
projected along the $x$-axis, whose ``backward-segments''
are of longer duration than the ``forward-segments''.
Similarly, when going over from the quasi-one-dimensional 
paths in Fig. 1 (zero ``width'')
to the corresponding meandering structures with a finite ``width'', 
we numerically verified (not shown) that the basic qualitative 
response behavior in Fig. 4 remains unchanged 
upon extending the width of the black lines in Fig. 3a
at least as long as the corners of the structure
indicated by the encircled numbers 1 and 4 in Fig. 3a did not yet
merge (corresponding to a maximal width of about $0.182\, l_2$).

\acknowledgements
This work was supported by Deutsche Forschungsgemeinschaft under 
SFB 613 and RE 1344/3-1 and by the ESF-program STOCHDYN.

%
%
\begin{figure}[h]
\epsfxsize=\columnwidth
\epsfbox{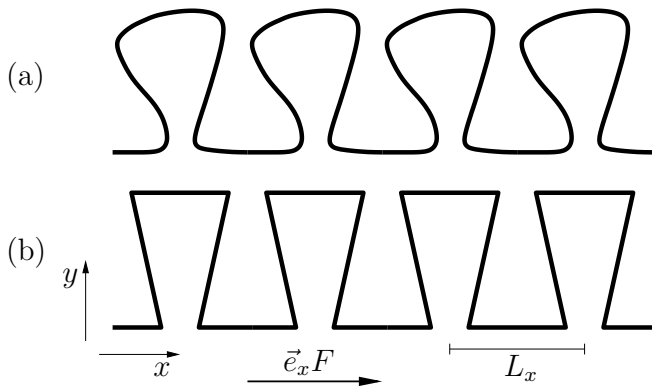}
\caption{
Examples of quasi-one-dimensional meandering 
paths in the $x$-$y$-plane (e.g. generated by a potential
which is zero in the black and infinity in the white 
domains).
They are required to be periodic (with period $L_x$) along the
$x$-axis and confined along the $y$-axis.
A further essential feature is the existence of
sharp bends such that a steady motion along the meandering
path results in a back-and-forth motion when projected 
along the $x$-axis.
In general, no further symmetry is required, as exemplified with
(a). 
The piecewise linear example (b) of high symmetry
serves to simplify the discussion of the basic physical
mechanisms.
A homogeneous, static bias force $F$ is applied along the 
$x$-direction. 
}
\label{fig1}
\end{figure}
%
\begin{figure}[h]
\epsfxsize=\columnwidth
\epsfbox{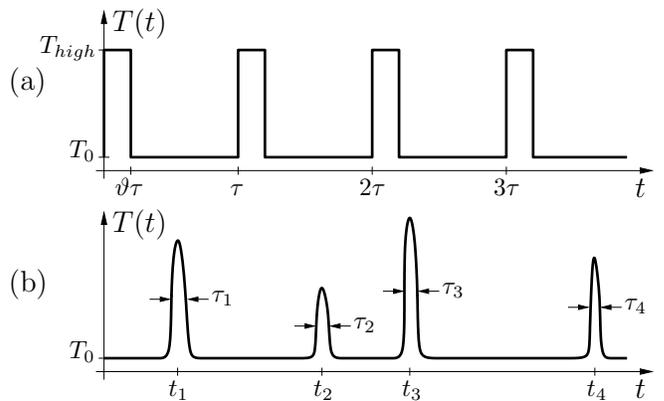}
\caption{
Examples of time-dependent temperatures $T(t)$.
(a) $T(t)$ is periodically switching between $T_{high}$ and $T_0$.
The total period is $\tau$ and the duration of the
high- and low-temperature segments $\vartheta\tau$ and
$(1-\vartheta)\tau$, respectively, with $\vartheta\in (0,1)$.
(b) $T(t)$ consist of a ``background'' $T_0$ and random 
``spikes'' at times $t_i$ of duration $\tau_i$.
Their average distance and duration are characterized 
according to Eqs. (\protect\ref{eq1b}), (\protect\ref{eq1c}).
}
\label{fig1neu}
\end{figure}
%
\begin{figure}[h]
\epsfxsize=\columnwidth
\epsfbox{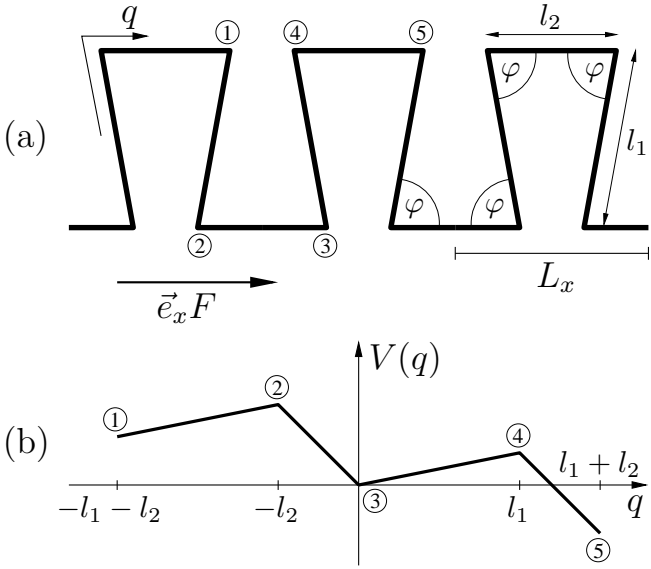}
\caption{(a) Same example as in Fig. 1b, parametrized by the
angle $\varphi$ and the lengths $l_1$, $l_2$.
(b) The corresponding one-dimensional potential $V(q)$ 
as described above Eq. (\protect\ref{eq2}), i.e.,
$V'(q) = -F$ for $q \in [-l_2,0]$ or $q\in[l_1,l_1+l_2]$
and $V'(q) = F \cos\varphi$ for $q\in [-l_1-l_2,-l_2]$
or $q \in [0,l_1]$.
The encircled numbers mark corresponding positions
in (a) and (b). 
}
\label{fig2}
\end{figure}
\begin{figure}[h]
\epsfxsize=\columnwidth
\epsfbox{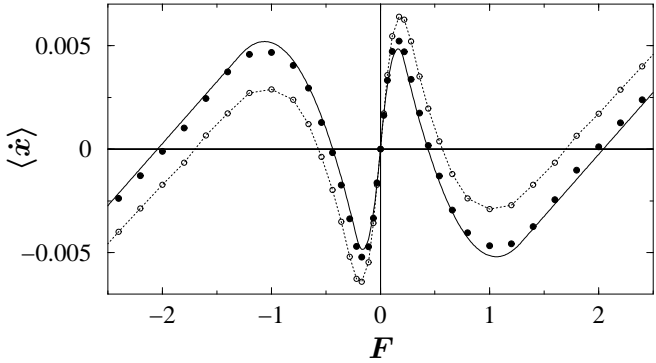}
\caption{
Average particle current (\protect\ref{eq0}) versus static force $F$
for the meandering path in Fig. 3a with $l_1=3$, $l_2=2$, 
$\varphi=75^o$, and a time-periodic temperature according to
Fig. 2a with $\tau=10$, and $\vartheta=0.1$.
Filled circles: $T_0=0.001$, $T_{high}=50$.
Open circles: $T_0=0.01$, $T_{high}=5$.
Shown are results from numerical simulations of 
(\protect\ref{eq1})-(\protect\ref{eq7}) in dimensionless units with 
$\eta=k_B=1$.
The numerical uncertainty is about the symbol size.
Solid line: The theoretical approximation
(\protect\ref{eq16}), coinciding with (\protect\ref{eq17}) since $T(t)$ is 
periodic.
Dotted line: guide for the eye.
}
\label{fig3}
\end{figure}
\begin{figure}[h]
\epsfxsize=\columnwidth
\epsfbox{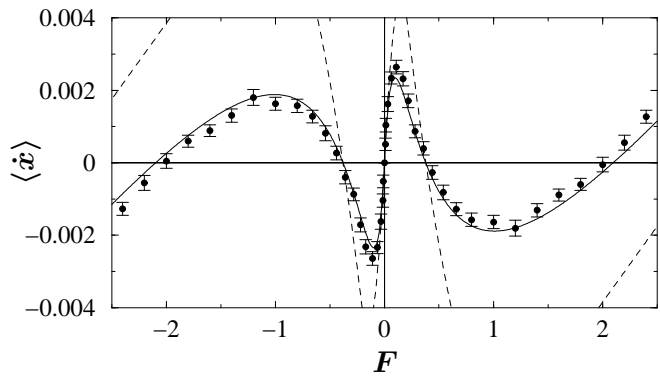}
\caption{Filled circles: Same as in Fig. 4 with the exception 
that the spiking times $t_i$ are not periodic [as in Fig. 2a]
but rather randomly sampled [as in Fig. 2b] 
according to a Poisson process with average 
interspike distance $\tau=10$ [cf. (\protect\ref{eq1b})].
The relative duration $\tau_i=0.07=\vartheta$ 
is the same for all spikes $i$ [cf. (\protect\ref{eq1c})].
Solid line: Theoretical approximation (\protect\ref{eq16}). 
Dashed line: Simplified approximation (\protect\ref{eq17}). 
}
\label{fig4}
\end{figure}

\end{document}